\documentclass[aps,pra,reprint,showpacs,superscriptaddress,longbibliography]{revtex4-2}
\usepackage{mathtools,amssymb,graphicx,units}

\usepackage[plainpages=false,pdfpagelabels,colorlinks=true,linkcolor=red,urlcolor=blue,citecolor=blue,pdftitle={Title},pdfauthor={},pdfdisplaydoctitle=true,pdfduplex=DuplexFlipLongEdge]{hyperref}
\usepackage{amsmath,graphicx,units}
\usepackage[table,xcdraw]{xcolor}
\usepackage{verbatim}
\usepackage[english]{babel}
\usepackage{color}

\newcommand{\tcr}{}

\newcommand{\up}{\uparrow}
\newcommand{\down}{\downarrow}

\begin{document}

\title{Dimensional crossover on multileg attractive-$U$ Hubbard ladders}

\author{Anastasia Potapova}
\author{Ian Pil\'{e}}
\affiliation{HSE University, 101000 Moscow, Russia}

\author{Tian-Cheng Yi}
\affiliation{Beijing Computational Science Research Center, Beijing 100084, China}

\author{Rubem Mondaini}
\affiliation{Beijing Computational Science Research Center, Beijing 100084, China}

\author{Evgeni Burovski}
\affiliation{HSE University, 101000 Moscow, Russia}

\date{\today}

\begin{abstract}
We study the ground state properties of a polarized two-component Fermi gas on multileg attractive-$U$ Hubbard ladders. Using exact diagonalization and density matrix renormalization group method simulations, we construct grand canonical phase diagrams for ladder widths of up to $W=5$ and varying perpendicular geometries, characterizing the quasi-one-dimensional regime of the dimensional crossover. We unveil a multicritical point marking the onset of partial polarization in those phase diagrams, a candidate regime of finite-momentum pairing.  We compare our findings with recent experimental and theoretical studies of quasi-one-dimensional polarized Fermi gases.
\end{abstract}

\maketitle

\section*{Introduction}

Quantum liquids in low dimensions are predicted to support a variety of exotic superfluid phases, the so-called Fulde-Ferrell-Larkin-Ovchinnikov state (FFLO)~\cite{Fulde1964, Larkin1964} being a prime example. Here the superconducting order parameter is modulated in space due to Cooper pairing at finite center-of-mass momentum. The FFLO state remains elusive despite being predicted more than fifty years ago.
There is a growing body of indirect evidence~\cite{* [{and references therein}]{Agterberg2020_cuprates_review}} for FFLO in experiments in $\mathrm{CeCoIn_5}$ in rotating magnetic field~\cite{Lin_rotating_field2020}, organic superconductors ~\cite{Wosnitza2018_organic1, Imajo_organic2}, pnictides~\cite{Cho2017} and cuprates~\cite{Du2020_cuprate1}. Yet, direct experimental observation of the FFLO state in condensed matter systems remains challenging. One reason the FFLO phase is elusive is that the impurity scattering can completely suppress the FFLO long-range ordering~\cite{Song2019_multilayer_fflo}.

Ultracold gases provide a promising platform for the quest for FFLO-like physics thanks to the ability to engineer clean systems and unprecedented control over polarization, interactions, and geometry. Several geometries have been explored: spherically symmetric three-dimensional (3D) traps \cite{Shin2006, Zwierlein2006, Zwierlein2006_2}, elongated cigar-shaped traps \cite{Partridge2006, Liao2010} and arrays of tubes \cite{Revelle2016}. Due to an applied trapping potential, the polarized gas cloud phase separates into the unpolarized (equal densities, EQ) and partially polarized (PP) regions---and the PP gas is a candidate for the FFLO state. The phase boundaries between regions can be directly measured via an \textit{in situ} imaging.
By varying either the confinement ratio or the intertube tunneling, experiments probe the 1D-to-3D crossover of polarized Fermi gases.
An immediate conclusion from the experiments is that the relative locations of the PP and EQ states in an external potential are inverted between the 1D and 3D limits: in 3D, the EQ phase occupies the center of the trap, while in 1D the center is PP and EQ is pushed to the wings.

The limiting cases of the 1D-to-3D crossover of a polarized two-component Fermi gas are relatively well understood from the theory side. In the 3D case, the uniform polarized state with FFLO long-range ordering is only stable in a narrow range of interactions and polarizations \cite{Sheehy2006}. In two dimensions (2D), the FFLO state is believed to be stable in a wide range of parameters \cite{Gukelberger2016}, including the regime of light doping from half-filling with small spin polarization \cite{Vitali2022}. Mean-field studies \cite{Datta2019}, and Monte Carlo simulations indicate further possibilities for competing long-range orders. For a review, see, e.g.  \cite{TormaReview2018} and references therein.

In 1D, the whole PP phase has a dominant FFLO-like algebraic ordering \cite{Yang2001} and is stable in a wide range of interactions. For a trapped 1D gas, the phase separation scenario is consistent with the experiments in highly elongated cigar-shaped traps \cite{Orso2007}. Following the initial field-theoretical \cite{Yang2001} and Bethe ansatz \cite{Orso2007} treatments, the 1D polarized Fermi gas was extensively studied numerically, using DMRG \cite{FHM2007_1D, Rizzi2008, Tezuka2008, Zwerger2010, oneleg_ladder} and QMC simulations \cite{Batrouni2008, Casula2008, Wolak2010}.

The 1D-to-3D crossover for polarized Fermi gases has been studied in the mean-field approximation (MFA) in Refs. \cite{Parish2007, Sun2013, Sundar2020}. While the MFA approach allows tracing the main features of the crossover qualitatively, it fails to capture several qualitative effects seen in experiments at low densities close to the 1D limits: the MFA misrepresents the topology of the grand canonical phase diagram in the 1D limit \cite{Parish2007, Sundar2020}.

This paper addresses strong correlation effects beyond MFA in the quasi-1D regime. To this end,  we study the quasi-1D attractive Hubbard model on wide ladders of up to five legs. We employ two numerically unbiased approaches: exact diagonalization (ED) of the Hamiltonian on small clusters of varying aspect ratio and density-matrix renormalization group (DMRG) simulations \cite{White1992, Schollwoeck2005} in ladder-geometries. Our numerical results for two and three legs are consistent with previous simulations of Refs.\  \cite{twoleg_ladder} and \cite{Burovski2019}. By considering wider ladders, we map out the evolution of the $T=0$ grand canonical phase diagram for quasi-1D geometries from a strict 1D limit to higher dimensions. Specifically, we find that (i) in presence of the external trapping potential, the 3D scenario of the phase separation is a robust feature that sets in immediately away from the strict 1D limit, and (ii) the stability region of the FFLO-like phase shrinks upon increasing the ladder's width, i.e., away from the strict 1D limit.

We also note that considering wide ladders (with flat or cylinder geometry in the perpendicular direction) is becoming feasible from the computational point of view and has been recently used for various problems related to variants of the repulsive Hubbard model \cite{Kantian2019, Chung2020, Huang2022}.
%

\section*{Model and Methods}
We consider the attractive-$U$ Hubbard model of a two-component Fermi gas at zero temperature on a $W \times L$ ladder defined by the Hamiltonian:

\begin{equation} \label{Hubbard}
\hat{H}=-t \sum_{\langle ij \rangle, \sigma} \left( \hat{c}_{i,\sigma}^{\dagger} \hat{c}_{j, \sigma}^{\phantom \dagger} + {\rm h.c.} \right) + U \sum_{i} \hat{n}_{i, \uparrow} \hat{n}_{i, \downarrow} \ .
\end{equation}
Here $\hat{c}_{i,\sigma}^{\dagger}$ ($\hat{c}_{i,\sigma}$) is a creation (annihilation) operator of a fermion with pseudospin $\sigma=\uparrow, \downarrow$ on site $i$; $\hat{n}_{i, \sigma} = \hat{c}_{i,\sigma}^{\dagger} \hat{c}_{i,\sigma}^{\phantom \dagger}$ is the corresponding number operator.
We set the hopping amplitude $t = 1$ as the energy scale, and $U < 0$ is the local Hubbard attraction parameter between two fermions with opposite spins. The summation over $i$ in Eq.~\eqref{Hubbard} runs over $L \times W$ sites, where $L$ is the length of an $W$-leg ladder.

In the canonical ensemble, the ground state energies, $E_0(N_\uparrow, N_\downarrow)$, of Eq.~\eqref{Hubbard} are characterized by the cardinality of spin-up and spin-down fermions, $N_\uparrow$ and $N_\downarrow$, or equivalently by the filling fractions, $n_\sigma= N_\sigma / LW$. Considering the case where $n_\uparrow \geqslant n_\downarrow$, the following ground states are of interest:
\begin{enumerate} 
    \item Vacuum (V): $n_\uparrow = n_\downarrow = 0$
    \item Equal densities (EQ): $n_\uparrow = n_\downarrow$. At $T=0$ this phase supports the BCS regime.
    \item Partially polarized (PP) phase: $n_\uparrow > n_\downarrow > 0$. This phase is the FFLO candidate.
    \item Fully polarized (FP) phase: $n_\uparrow > 0$ and $n_\downarrow = 0$.
\end{enumerate}
For a lattice model~\eqref{Hubbard}, we further distinguish whether the majority spin band is filled ($n_\uparrow = 1$---we call this the FP$_2$ phase) or not (i.e., $n_\uparrow < 1$---we call this the FP$_1$ phase). 

Changing the variables from the canonical to the grand canonical ensemble, we introduce the effective magnetic field $h$ and the chemical potential $\mu$ via
\begin{equation} \label{eq:mu_main_text}
\mu = {\left( \frac{\partial E_0}{\partial N} \right)}_{P} \;,\qquad
h = {\left( \frac{\partial E_0}{\partial P} \right)}_{N} \;,
\end{equation}
where $N = N_{\uparrow} + N_{\downarrow}$ is the total particle number and $P = N_{\uparrow} - N_{\downarrow}$ is the polarization.

The phase boundaries are found by approximating the derivatives \eqref{eq:mu_main_text}  with finite differences --- this is equivalent to comparing the ground state energies of different phases. For instance, the boundary between V and EQ is given by comparing the energy of a two-particle state with the zero energy of an empty band, i.e. $\mu = E_0(1, 1) / 2$. Likewise, the boundary between FP$_1$ and V is related to the bottom of the single-particle band, $\mu = -h + E_0(1, 0)$. Further details, including the finite-difference approximations to Eq.\ \eqref{eq:mu_main_text} for computing the EQ-PP and FP-PP phase boundaries, are given in 
Appendix~\ref{app:appendix_a}


\section*{Results and discussion}

\tcr{We resort to} DMRG computations~\footnote{For that, we make use of both ALPS \cite{ALPS2} and iTensor \cite{iTensor} implementations.} to extract the ground-state energies $E_0(N_\uparrow, N_\downarrow)$ of the Hamiltonian \eqref{Hubbard} on ladders of length $L$ and widths $W$ of up to $W=5$ with open boundary conditions in \emph{both} 
dimensions. For that, we fix the length $L=40$, and check that using larger values of $L$ does not significantly change the results. We also run simulations of $W=3, 4, 5$ with the cylinder geometry (i.e., periodic boundary conditions in the perpendicular direction). \tcr{Appendix ~\ref{app:EQ} highlights the necessity of using DMRG techniques by showing that ED results are largely impacted by finite-size effects in the small clusters amenable to computations.} For the interaction strength, we take $U = -7$. At these parameters, we typically use up to 2000 states for the bond dimension, so that the truncation error in the DMRG process is below $10^{-8}$ for $W \leqslant 3$ and below $10^{-6}$ for $W=4, 5$. 

\begin{figure}[hbt]
\includegraphics[width=0.99\columnwidth]{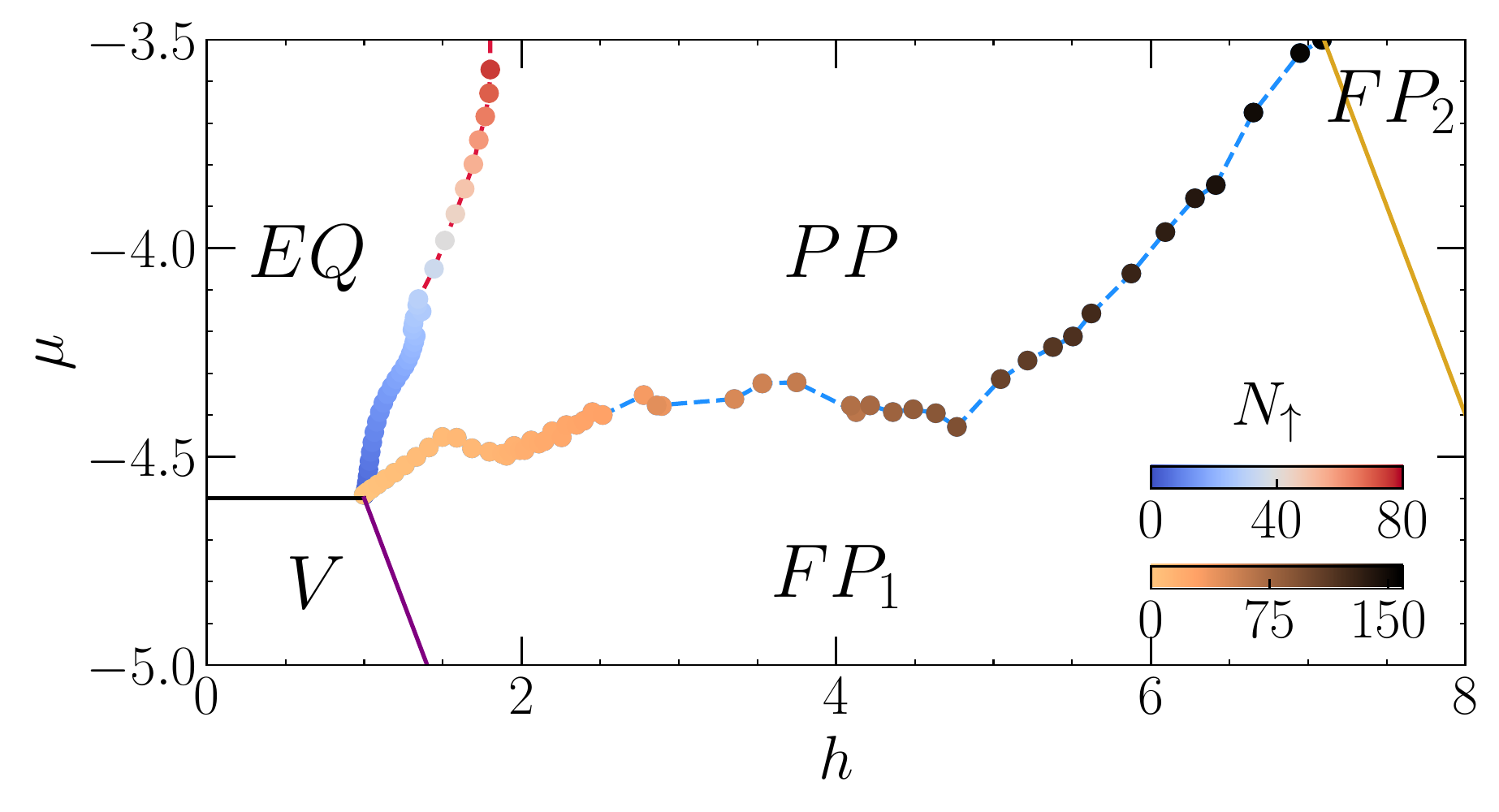}
\caption{The phase diagram for $W = 4$, $U = -7$, and $L = 40$. \tcr{Phases considered are vacuum (V) ($n_\uparrow = n_\downarrow = 0$), equal densities (EQ)($n_\uparrow = n_\downarrow$), partially polarized (PP)($n_\uparrow > n_\downarrow > 0$) and two fully polarized phases: FP$_1$ with $n_\uparrow > 0, n_\downarrow = 0$ and FP$_2$ with $n_\uparrow = 0$ , $n_\downarrow > 0$. Here we only show the region $h > 0$ and $\mu < U/2$ because the phase diagram is symmetric for $h\to -h$ and $\mu\to U/2 - \mu$.} Filled circles are DMRG results, and dashed lines connecting them are guides to the eye. \tcr{The marker colors are mapped by the value $N_\uparrow$ that identifies the boundary line according to the corresponding colorbars -- See Eqs.~\eqref{eq:EQ-PP_mu} and \eqref{eq:PP-FP1_mu}.}
\label{fig:w4}}
\end{figure}

For the strictly 1D model ($W=1$), our numerical results are consistent with previous Bethe Ansatz results \cite{Orso2007}, and DMRG simulations \cite{oneleg_ladder}. For $W=2$ and $W=3$ our results agree with previous DMRG simulations \cite{Burovski2019, twoleg_ladder}.
Figure\ \ref{fig:w4} shows the grand canonical phase diagram for the $W=4$ ladder---other $W>2$ diagrams are qualitatively similar. We also checked that imposing periodic boundary conditions in perpendicular directions does not significantly alter the results. 
Several features stand out in Fig.\ \ref{fig:w4}. First, the PP phase---the candidate to support the FFLO-like physics---occupies a significant area on the phase diagram. This fact was previously seen in simulations of $W \leqslant 3$ ladders, and our results for $W=4, 5$ support this being a robust feature of quasi-1D geometries.

\begin{figure}[htb]

\includegraphics[width=0.99\columnwidth]{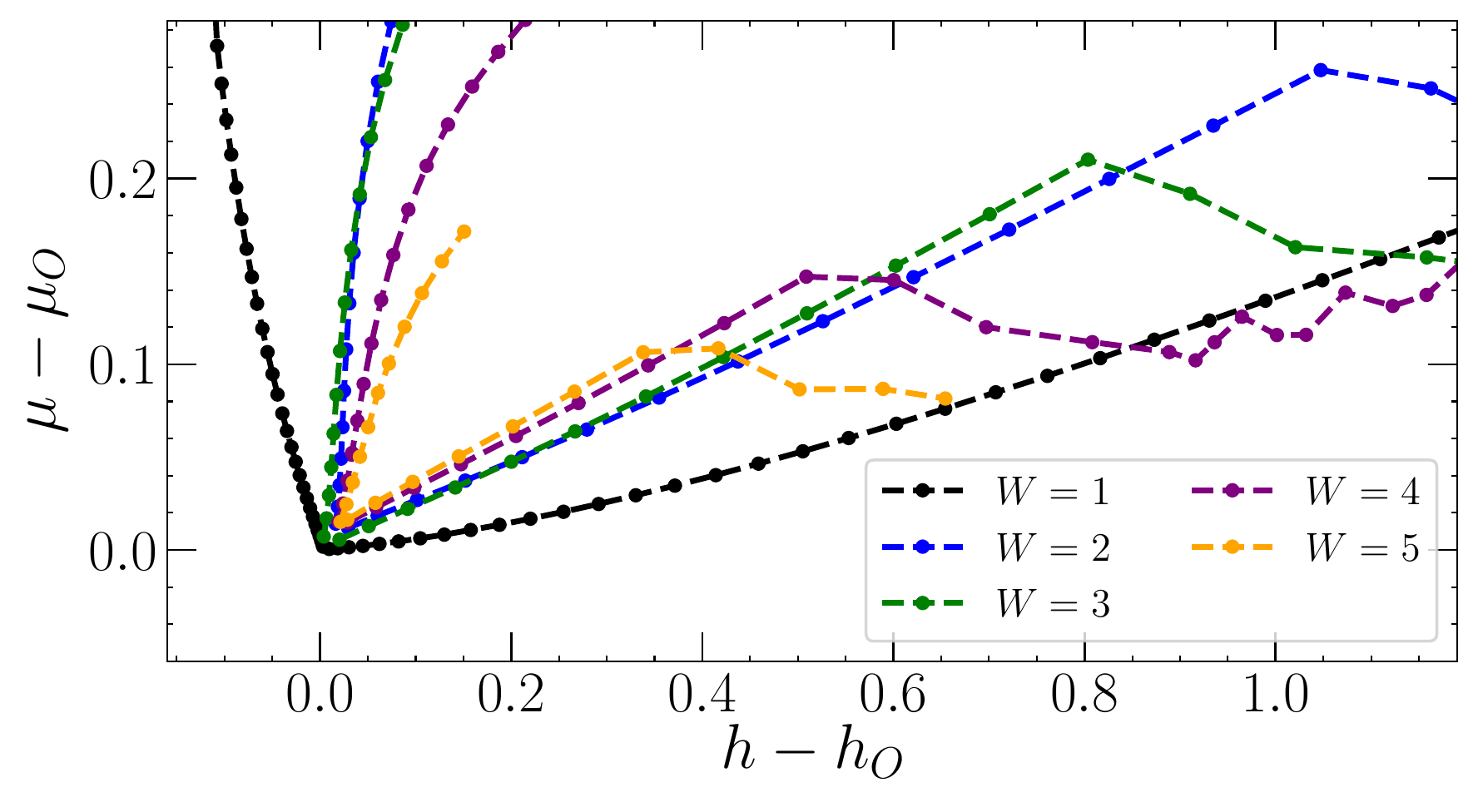}
\caption{The vicinity of the ``multicritical'' point $O$ for $W = 1$ (black line), for $W = 2$ (blue line), $W = 3$ (green line), $W = 4$ (purple line) and $W = 5$ (yellow line) and $L=40$. 
Here we shift the boundaries for each $W$ value so that for each $W$'s ``multicritical'' point $O$ is at the origin. We also only show for each $W$ the  EQ-PP and PP-FP$_1$ lines. Filled symbols are DMRG results, and dashed lines are to guide an eye. We also note that the vicinity of the multicritical point corresponds to the continuum limit of the lattice model \eqref{Hubbard} since the filling fractions are small, $N_\sigma \ll WL$.
}
\label{fig:comp12345}
\end{figure}

Second, the EQ, PP and FP$_1$ phases meet at a ``multicritical'' point $O$, whose location is fixed by the few-body physics of the model, $(\mu_O, h_O) = (E_0(1, 1)/2, -E_0(1, 1)/2 + E_0(1, 0))$. That the EQ-FP phase boundary also terminates at the point $O$ can be traced to the fact that the only bound states in the Hubbard model \eqref{Hubbard} are pairs of spin-up and spin-down fermions \cite{Orso2010trimers}. We checked that the binding energies of three-body states (``trimers'') are zero for all values of $W$, modulo finite-$L$ corrections.
We also note that one of the main artifacts of the mean-field approximation is the prediction that the point $O$ splits into two tricritical points \cite{Parish2007, Sundar2020}. A compilation of parts of the phase diagram close to the multicritical point $O$ with a growing number of legs is shown in Fig.~\ref{fig:comp12345}.

\textit{The PP-FP$_{1}$ (``polaron'') line.---} The phase boundary between FP$_1$ and PP phases corresponds to the so-called Fermi-polaron problem: a single spin-down particle in a sea of $N_\uparrow$ spin-up fermions (the points along the line correspond to the increasing $N_\uparrow$). While for $W=1$ the phase boundary is smooth and monotonic, for $W>1$ the boundary has kinks, which can be traced to the filling of non-interacting energy bands in the spectrum \cite{twoleg_ladder}. In general, for a $W$-leg ladder, there are $W$ (overlapping) branches of the single-particle spectrum, and kinks on the ``polaron'' line in the $\mu$-$h$ plane correspond to onsets of partial filling of multiple branches. Whenever multiple branches are partially filled, the ``polaron'' line is non-monotonic and has additional chaotic oscillations. These are likely due to the finite-$L$ level spacing of the motion along the ladder.

Another point to note is that for $W \geqslant 2$, the initial part of the ``polaron'' line---which corresponds to the filling of the lowest single-particle band--- is close to linear. This initial linear behavior is followed by a cusp and a sharp downturn for the range where the second branch starts filling. This behavior was first observed on a two-leg ladder in Ref.\ \cite{twoleg_ladder} and our simulations indicate that it persists for larger values of $W$ and varying the boundary conditions in the perpendicular direction.

\emph{The EQ-PP line.---} We now turn our attention to the boundary between the EQ and PP phases. The main difference between the strictly 1D case of $W=1$ and $W> 1$ is that in the vicinity of the ``multicritical'' point $O$, the slope of the boundary $\partial h / \partial \mu < 0$ for $W=1$ \cite{Orso2007, oneleg_ladder} while for $W>1$ the slope of the boundary has the opposite sign, $\partial h / \partial \mu > 0$. This has been first noted for $W=2$ in Ref. \cite{twoleg_ladder}, and later confirmed for $W=2, 3$ in Ref. \cite{Burovski2019}. Our results, illustrated in Figs.\ \ref{fig:w4} and \ref{fig:comp12345} for up to $W=5$ corroborate the conclusion that the sign change is a robust feature that differentiates the strict 1D limit from quasi-1D geometries with $W \neq 1$.

In the presence of an external trapping potential, the shell structure \tcr{(or phase separation)} of the atomic cloud can be directly read off the grand canonical phase diagram in the local density approximation. For a spin-independent trapping potential, $V(x)$, applied along the ladder direction, the cloud structure corresponds to $\mu\to \mu - V(x)$, i.e., a vertical cut on the $h-\mu$ plane. This way, the sign of the slope of the EQ-PP boundary differentiates between the 1D behavior with a two-shell structure where the PP phase occupies the center of the trap \cite{Liao2010, Revelle2016}, and a higher-dimensional behavior where the shell structure is inverted: the center of the trap has equal densities and the PP phase occupies an outer shell  \cite{Shin2006, Zwierlein2006, Revelle2016}.
Our results, illustrated in Fig.\ \ref{fig:w4} and \ref{fig:comp12345}, show that the 3D-like shell structure is a robust feature of the quasi-1D geometries. 
We stress that in our simulations, the EQ-PP boundary is monotonic, in contrast to the mean-field results \cite{Parish2007, Sundar2020}, where the MF approximation generates artifacts at low density (equivalently, strong interactions) which lead to an apparently reentrant behavior \footnote{In fact, we do observe some reentrant behavior if we allow the hopping amplitudes along the rungs of the ladder, $t_\perp$ to differ from the hopping amplitudes along the ladder, $t_\parallel$. We reserve a thorough discussion of these phenomena for a separate investigation.}.

\emph{The PP phase stability region.---}%
In Fig.\ \ref{fig:comp12345}, we compare the low-filling parts of phase diagrams for ladders with $W=1$ to $W=5$, where we shift the values of $\mu$ and $h$ for each $W$ so that their respective points $O$ coincide. In Fig.\ \ref{fig:comp12345} we only show the EQ-PP and PP-FP$_1$ lines for clarity. 
It is immediately seen from Fig.\ \ref{fig:comp12345} that with increasing the width $W$, the EQ-FP boundary shifts to the right while the ``polaron'' line shifts upwards. 

We thus conclude that the stability region for the PP phase---the candidate phase for the FFLO-like physics---shrinks on approach away from a strict 1D limit \emph{while in the quasi-1D regime}. The last comment is due to the fact that the position of the kink and a downturn on the ``polaron'' line, visible in Fig.\ \ref{fig:comp12345}, shifts to smaller values of filling fraction, $N_\uparrow/WL$, for increasing $W$. Since the position of the kink is related to the filling of the second single-particle branch of the transversal motion, and the gap between branches goes to zero in 2D or 3D, our approach does not allow us to make definite statements about the behavior of the stability region of the FFLO phase for $W\sim L$.

\emph{The FLLO in the PP phase.---} To probe the FFLO character of the PP phase, we compute the superconducting correlation functions, $\Gamma(\mathbf{x}, \mathbf{x}_0) = \langle \hat \Delta^\dagger(\mathbf{x}_0) \hat \Delta(\mathbf{x} + \mathbf{x}_0)\rangle$, where $\hat \Delta(\mathbf{x}) = \hat c_{\mathbf{x}\up} \hat c_{\mathbf{x}\down}$ annihilates a pair of fermions at a lattice site with coordinates $\mathbf{x}$. In the 1D limit, $W=1$,  $\Gamma$ is expected \cite{Yang2001} to decay algebraically with $|\mathbf{x}|$ and oscillate with the typical FFLO momentum $Q = \pi (n_\uparrow - n_\downarrow)$. 

\begin{figure}[thb]
\includegraphics[width=0.99\columnwidth]{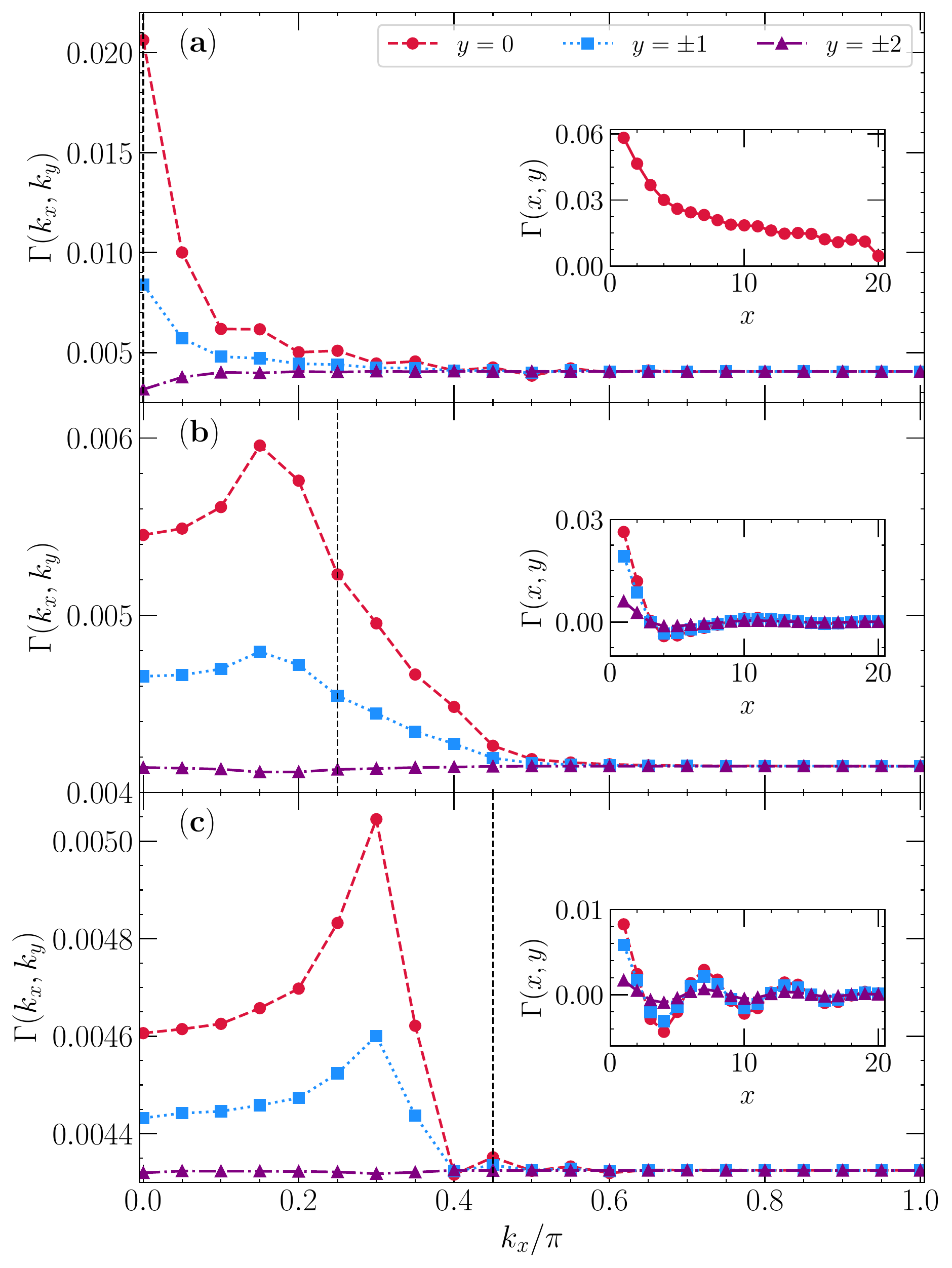}
\caption{Fourier-transformed pair-pair correlation functions $\widetilde\Gamma(k_x, k_y)$, in momentum space on a $5\times 40$ ladder; insets show the corresponding real-space decay of such correlations. The total density $n = n_\uparrow+ n_\downarrow = 0.1$ is fixed, whereas the polarization changes. In (a) $N_\uparrow=N_\downarrow = 10$; (b) $N_\uparrow = 15$, $N_\downarrow = 5$; and (c) $N_\uparrow = 19$, $N_\downarrow=1$. The vertical dashed lines depict the  characteristic momenta from FFLO in one-dimensional systems, $Q = \pi (N_\uparrow - N_\downarrow ) / L$. See text for discussion. The different curves in the main panels show the different allowed transverse momentum modes $k_y$, whereas in the insets the actual transversal coordinate $y$.
\label{fig:pair-pair-corr}}
\end{figure}

To trace the behavior of the superconducting correlations in ladder geometries, we consider a $W=5$ ladder and take $\mathbf{x}_0$ at the center of the lattice, $\mathbf{x}_0 = (L/2, 0)$, where the legs of the ladder are indexed from $-(W-1)/2$ to $(W-1)/2$. Figure~\ref{fig:pair-pair-corr} shows the Fourier transform, $\widetilde\Gamma(k_x, k_y)$, of $\Gamma(\mathbf{x}, \mathbf{x}_0)$ for $L=40$, and different polarizations $(N_\uparrow, N_\downarrow) = (10, 10)$ [panel (a)], $(15, 5)$ [panel (b)], and $(19, 1)$ [panel (c)], with a fixed total occupancy. The insets display the corresponding real-space correlations along the ladder, resolved by the $y$-coordinate in the transverse direction. 

For balanced occupancies, the real-space pair-pair correlations decay monotonically with distance which translates to a $\widetilde\Gamma(k_x, k_y)$ peaked at zero-momentum [Fig.~\ref{fig:pair-pair-corr}(a)]. In turn, once a finite density-imbalance is selected, oscillations on the correlations in real space are clearly visible, resulting in peaks of the Fourier-transformed correlations at finite momentum [Figs.~\ref{fig:pair-pair-corr}(b) and (c)]. This is direct evidence of FFLO-type physics occurring in wider ladders. The peak location in $k_x$, however, which occurs at the same point irrespective of the $k_y$ value, is not aligned with the expected typical FFLO momentum of 1D systems. In practice, a combination of modes sets in, a fact already appreciated in $W=2$ ladders in Ref.~\cite{twoleg_ladder}.

\begin{figure}[thb]
\includegraphics[width=0.99\columnwidth]{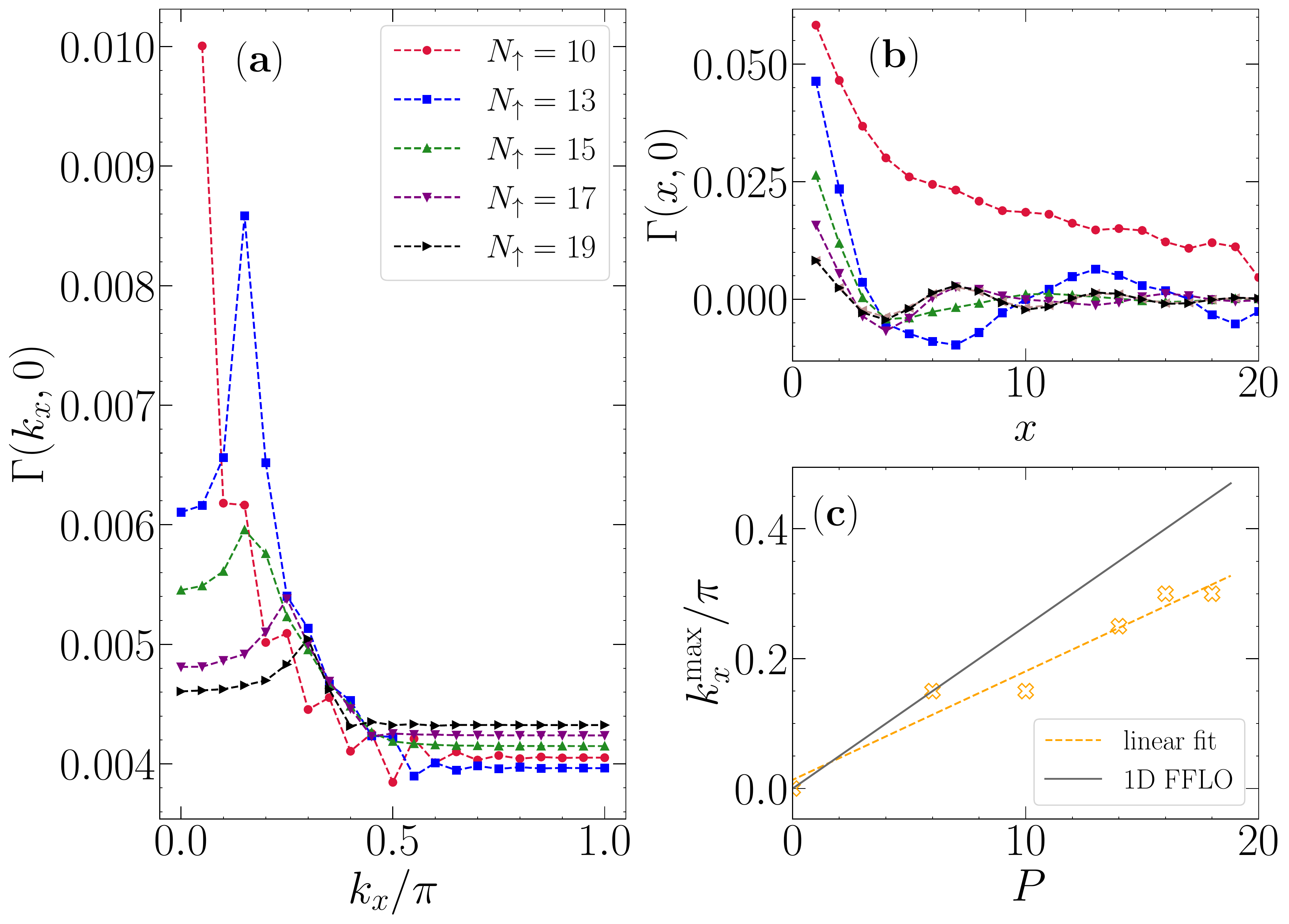}
\caption{(a) Fourier transformed pair-pair correlations $\widetilde\Gamma(k_x, k_y)$ with zero-transverse momentum $(k_y=0)$, and fixed particle-number $N=20$. Data at $k_x=0$ and $P=0$ is omitted for clarity. (b) The corresponding real-space correlations along the central leg ($y=0$) of the $5\times 40$ ladder. (c) The compilation of the peak-locations in panel (a) vs.~the polarization $P=N_\uparrow-N_\downarrow$ accompanied by a linear fitting and the typical momentum dependence in 1D FFLO problems.} 
\label{fig:fig5}
\end{figure}

Since this finite-momentum peak depends on the total polarization, we compile in Fig.~\ref{fig:fig5} the Fourier transformed $\widetilde\Gamma(k_x, 0)$ correlations with growing polarization $P=0\ldots 18$, and fixed particle number $N=20$ on the same $5\times40$ ladder. A finite-momentum peak at $k_x^{\rm max}$ occurs only when $P>0$ [Fig.~\ref{fig:fig5}(a)], tied to the appearance of oscillations with the distance of the real-space correlations [Fig.~\ref{fig:fig5}(b)]. This peak position linearly follows the polarization $P$, as with the expected typical FFLO momentum $Q = \pi(n_\uparrow - n_\downarrow)$, albeit with a slightly decreased slope [Fig.~\ref{fig:fig5}(c)]. 

Whether such oscillations and finite-momentum pairing remain in truly two-dimensional systems, studied utilizing unbiased methods, is an open question. Our results point out that still in the quasi-one-dimensional limit, they yet occur. Lastly, it is important to emphasize that mean-field investigations unveil the possible scenarios of normal-polarized phases~\cite{Parish2007, Sundar2020}, that is, without pairing formation. In studies that employed unbiased methods as ours in $W=2$ ladders~\cite{twoleg_ladder}, such regimes have been ruled out but deserve future investigation at wider ladders.

\section*{Summary}
We study the quasi-1D limit of the dimensional crossover of a polarized Fermi gas. By simulating the attractive-$U$ Hubbard model on wide ladders encompassing up to $W=5$ legs, we trace the evolution with $W$ of the grand canonical phase diagram, which features BCS-like and FFLO-like states. Our simulations complement earlier mean-field-based investigations of the dimensional crossover, where the uncontrollable nature of the mean-field approximation leads to artifacts at strong coupling for quasi-1D systems. We show that a qualitative difference in a shell structure of a gas in an external potential, observed in experiments with ultracold gases, is a robust feature that differentiates a strict 1D limit and higher dimensions, including quasi-1D systems. Our simulations indicate that the stability region of the FFLO phase in the quasi-1D regime shrinks with increasing the ladder width away from the strict 1D case. 

\begin{acknowledgments}
We heartily thank Th.~Jolicoeur and G.~Misguich for multiple fruitful discussions. We acknowledge financial support by RFBR and NSFC, project number 21-57-53014. R.M.~acknowledges support from the NSFC Grants No.~U2230402, 12111530010, 12222401, and No.~11974039. Numerical simulations were carried out using the HSE HPC resources~\cite{Kostenetskiy2021}, and the Tianhe-2JK at the Beijing Computational Science Research Center.
\end{acknowledgments}

\appendix

\section{Extracting phase boundaries in ED and DMRG data}
\label{app:appendix_a}
In the main text, we argue that an analysis of the energetics of different filling sectors of the Hamiltonian can extract the different phases on the $\mu-h$ plane. Here we present this analysis in an expanded level of detail. In the canonical ensemble, the ground state energies, $E_0(N_\up, N_\down)$, of Eq.~\eqref{Hubbard} are characterized by the numbers of spin-up and spin-down fermions, $N_\up$, and $N_\down$. Considering the case where $N_\up \geqslant N_\down$, the following ground states are of interest:

\begin{enumerate} 
    \item Vacuum (V): $N_{\uparrow} = N_{\downarrow} = 0$
    \item Equal Densities (EQ): $N_{\uparrow} = N_{\downarrow}$. At $T=0$ this phase supports the BCS ordering.
    \item Partially polarized (PP) phase: $N_{\uparrow} > N_{\downarrow} > 0$. This phase is the FFLO candidate.
    \item Fully polarized (FP) phase: $N_{\uparrow} > 0$ and $N_{\downarrow} = 0$.
\end{enumerate}
For a lattice model, we further distinguish whether the majority spin band is filled ($N_\up = L\times W$---we call this the FP$_2$ phase) or not (i.e., $N_\up < L\times W$---we call this the FP$_1$ phase). Changing the variables from the canonical to the grand canonical ensemble, we introduce the effective magnetic field $h$ and the chemical potential $\mu$ via
\begin{equation} \label{eq:mu_appendix}
\mu = {\left( \frac{\partial E_0}{\partial N} \right)}_{P} \;,\qquad
h = {\left( \frac{\partial E_0}{\partial P} \right)}_{N} \;,
\end{equation}
where $N = N_{\uparrow} + N_{\downarrow}$ is the total particle number and $P = N_{\uparrow} - N_{\downarrow}$ is the polarization.

The phase boundaries are found by approximating the derivatives \eqref{eq:mu_appendix}  with finite differences, leading to

\begin{enumerate}
\item The boundary between EQ and PP:
\begin{equation} \label{eq:EQ-PP_mu}
\mu, h = \frac{E_0(N_{\uparrow} + 1, N_{\downarrow} \pm 1) - E_0(N_{\uparrow} ,N_{\downarrow})}{2}
\end{equation}

where on the right-hand side the plus sign gives $\mu$ and the minus sign gives $h$. Note that the phase boundary corresponds to $N_\up = N_\down$ in Eq.\ \eqref{eq:EQ-PP_mu}.


\item Likewise, the PP-FP$_1$ boundary between  is given by
\begin{equation} \label{eq:PP-FP1_mu}
\mu, h = \frac{E_0(N_{\uparrow} \pm 1, 1) - E_0(N_{\uparrow}, 0)}{\pm 2}
\end{equation}


\item The boundary between EQ and V
\begin{equation}
\mu = \frac{E_0(1, 1)}{2}
\label{eq:V-EQ}
\end{equation}
     	
\item The boundary between V and FP$_1$
\begin{equation} \label{eq:V-FP}
\mu = - h + E_0(1, 0)
\end{equation}
     	
\item The boundary between FP$_1$ and FP$_2$

\begin{equation} \label{eq:FP-FP}
\mu = - h - E_0(1, 0)
\end{equation}

\end{enumerate}

From the ED or DMRG data, we construct the grand canonical phase diagrams using Eqs.~\eqref{eq:EQ-PP_mu}-\eqref{eq:FP-FP} (where the phase boundaries are parameterized by $N_\uparrow$). For the strictly 1D model ($W=1$), our results are consistent with previous Bethe Ansatz results \cite{Orso2007}, and DMRG simulations \cite{oneleg_ladder}. For $W=2$ and $W=3$ our results agree with previous DMRG simulations \cite{Burovski2019, twoleg_ladder}.

\section{\tcr{ED results and comparison to global minimization}}
\label{app:EQ}

\begin{figure}
\includegraphics[width=0.99\columnwidth]{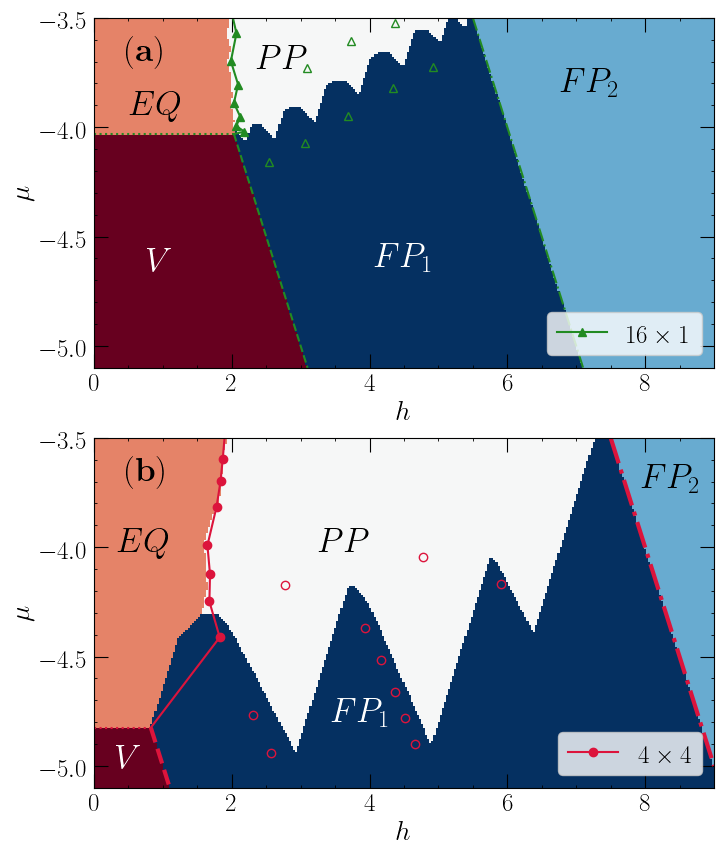}
\label{fig:ED_app}
\caption{Similar to the main text, the ground phase diagrams of the model \eqref{Hubbard} with $U = -7$ on 16-sites cluster with aspect ratios $l_x \times l_y$: (a) $16\times 1$ and (b) $4\times 4$. Phases are labeled similarly to Fig.~\ref{fig:w4}. Filled markers give the loci of the EQ-PP transition, according to Eq.~\eqref{eq:EQ-PP_mu}; empty markers the PP-FP$_1$ phase transition, Eq.~\eqref{eq:PP-FP1_mu}. The latter display substantially large oscillations stemming from finite-size effects. Dotted lines give the EQ-V transition; dashed, the V-FP$_1$ transition, and dash-dotted ones, the FP$_1$-FP$_2$ transition for the corresponding system sizes. Overlaid to it are colors that map the ground state according to the different phases when inspecting \textit{all} $(N_\uparrow, N_\downarrow)$ sectors of the Hamiltonian.}
\end{figure}

\tcr{We benchmark our exploration in the main text by using the ED method in small clusters. In particular, we contrast one- and two-dimensional geometries with the same number of sites (16-site clusters), observing how the dimensionality affects the $\mu-h$ phase diagram. While a characterization via finite differences is sufficient to reliably extract the phase boundaries as shown in the main text and further described in Appendix~\ref{app:appendix_a}, the small number of sites that can be tackled within this technique makes this procedure yield only a rough location of the limits between different phases.} 

\tcr{This is exemplified in Fig.~\ref{fig:ED_app} for the two aforementioned cluster shapes. An overall similar structure of the phase diagram is seen for the DMRG results in $W=4$ ladders [Fig.~\ref{fig:w4}], but the large oscillations on the phase boundaries given by the markers here attest to the large finite-size effects in such lattice sizes. Because all different sectors $(N_\uparrow, N_\downarrow)$ are readily available in this method, we can also draw the different boundaries by direct comparison of the total ground-state energy:}
\begin{eqnarray}
    E_0(\mu, h) = \min_{\{N_\uparrow, N_\downarrow\}}[&&E_0(N_\uparrow, N_\downarrow) \nonumber \\ &&- \mu(N_\uparrow + N_\downarrow) - h(N_\uparrow - N_\downarrow)]\ ,
    \label{eq:minimization}
\end{eqnarray}
\tcr{where direct inspection of the $E_0$'s can be used to classify which phase a given point $(\mu, h)$ the ground-state belongs to. By mapping the corresponding five different phases (see main text) to colors, we can then describe a refined phase diagram in Fig.~\ref{fig:ED_app}, which follows in tandem with the description of the finite difference estimation used in the main text. While also possible in the DMRG method, such exploration does not give a much better boundary between the phases because the finite-size effects are smaller in the larger cluster sizes we can tackle. Therefore, the finite difference scheme suffices in drawing the phase diagram there.}

\tcr{Finally, we emphasize that the finite differences procedure described in Appendix \ref{app:appendix_a} is numerically friendlier than the one given by Eq.~\eqref{eq:minimization} since only a limited set of ground-state energy computations is sufficient to obtain the phase boundaries. That is, other than the trivial cases, $E_0(1, 1)$ and $E_0(1, 0)$ (giving EQ-V, V-FP$_1$ and FP$_1$-FP$_2$ boundaries), the remaining boundaries require $E_0(N_\uparrow+1,N_\uparrow\pm1)$ and  $E_0(N_\uparrow,N_\uparrow)$ (EQ-PP); and  $E_0(N_\uparrow\pm1,1)$,  $E_0(N_\uparrow,0)$ for the PP-FP$_1$ transition. Thus with a limited number of runs, we can satisfactorily build the grand-canonical phase diagram of the model as shown in Fig.~\ref{fig:w4}, for example.}


\bibliography{references}
\end{document}